\documentstyle[seceq,preprint]{ptptex}
\newcommand{\ket }{\rangle }
\newcommand{\bra }{\langle }

\newcommand{\no }{\nonumber}

\preprintnumber[3cm]{%<-- [..]: optional width of preprint # column.
NUP-A-2000-9\\ June 2000}
\title{%        %You can use \\ for explicit line-break
How do Neutrinos Propagate ?
}
\subtitle{Wave-Packet Treatment of Neutrino Oscillation}    %use this when you want a subtitle

\author{%       %Use \sc for the family name
Yoshihiro {\sc Takeuchi},\footnote{E-mail address: yytake@phys.cst.nihon-u.ac.jp}  
 Yuichi {\sc Tazaki}, S.~Y.~{\sc Tsai},\footnote{E-mail address: tsai@phys.cst.nihon-u.ac.jp} \\ 
 and Takashi {\sc Yamazaki}
}

\inst{%         %Affiliation, neglected when [addenda] or [errata]
College of Science and Technology, Nihon University, \\
Kanda-Surugadai, Chiyoda-ku, Tokyo 101-8308, Japan
}

\recdate{%      %Editorial Office will fill in this.
}

\abst{%       %this abstract is neglected when [addenda] or [errata]
The wave-packet treatment of neutrino oscillation developed previously is extended to the case in which momentum distribution functions are taken to be 
a Gaussian form with both central values and dispersions depending on the mass eigenstates of the neutrinos. It is shown among other things that the velocity of the neutrino wave packets does not in general agree with what one would expect classically and that relativistic neutrinos emitted from pions nevertheless 
do follow, to a good approximation, the classical trajectory.
}

\begin{document}

\maketitle

\section{Introduction}

In a previous publication,\cite{rf:1} \ we have developed a wave-packet treatment of 
neutrino oscillation\cite{rf:2} for each of the "equal-energy prescription" and "equal-momentum prescription" (see Sect.3 below), and, by invoking relativistic kinematics as well, have derived the necessary conditions for oscillation to occur, 
which appear to have a form more well-defined and quantitative than what have 
been noted before.\cite{rf:3}\cite{rf:4}\cite{rf:5} \ The wave packets corresponding to each of the mass 
eigenstates of neutrinos in the case of the equal-energy (-momentum) 
prescription have been constructed from momentum (energy) distribution functions of a step-function type having a common central value and a common dispersion.

In the present note, we like to extend our wave-packet treatment of neutrino 
oscillation by introducing momentum distribution functions with both central 
values and dispersions depending on the mass eigenstates of the 
neutrinos and by replacing step-functions by Gaussian functions. Such extensions enable our treatment to have closer contact with actual experimental or observational situations, though our emphasis is still placed more or less on conceptual aspects of neutrino oscillation. We shall see among other things that the velocity of the neutrino wave packets does not in general agree with what one would expect classically and that relativistic neutrinos nevertheless do
follow, to a good approximation, the classical trajectory.\cite{rf:6} \ We shall also argue that the equal-energy (-momentum) prescription seems appropriate to approximately describe oscillation experiments which involve something corresponding to time- (space-) integration or averaging.

\section{Wave-packet treatment}

Let $|\nu_{\alpha}\ket$ ($\alpha = e, \mu, \tau$, etc.) be neutrinos associated with electron, muon, $\tau$ lepton, etc., which are mutually-orthogonal superpositions of the mass eigenstates $|\nu_i \ket $ having mass $m_i$ ($i$ = 1, 2, 3, etc.): 
%%%(2.1)
\begin{equation}
|\nu_{\alpha}\ket =\sum_i U_{\alpha i}|\nu_i\ket ~.
\end{equation}
Suppose a neutrino of flavor $\alpha$ is born at $t=0$ and propagates towards the $x$-direction. Then, its state vector at $x$ and $t$ may be written as\cite{rf:1}\cite{rf:4}
%%%(2.2)
\begin{equation}
|\nu_{\alpha}(x,t)\ket =\sum_iU_{\alpha i}g_i(x,t)|\nu_i\ket ~,
\end{equation}
%%%(2.3)
\begin{equation}
g_i(x,t)=\sqrt{\frac{1}{2\pi}}\int_{-\infty}^{\infty}dp_i f_i(p_i)e^{i(p_ix-E_it)}~,
\end{equation}
where $f_i(p_i)$ is the amplitude for creation of $|\nu_i\ket$ with momentum $p_i$ and energy $E_i = E_i(p_i) = \sqrt{p_i^2+m_i^2}$, normalized as\footnote{See Appendix A for some related remarks.}
%%%(2.4)
\begin{equation}
\int_{-\infty}^{\infty}dp_i|f_i(p_i)|^2 = 1~.
\end{equation}
The probability to find a neutrino with flavor $\alpha'$ at $x$ and $t$ is calculated as
%%%(2.5)
\begin{eqnarray}
P_{\alpha \to \alpha'}(x,t) &=& |\bra \nu_{\alpha'}|\nu_{\alpha}(x,t)\ket |^2 \no \\
&=& \sum_{i,j}U_{\alpha i}U_{\alpha' i}^*U_{\alpha j}^*U_{\alpha 'j}G_{ij}(x,t)~,
\end{eqnarray}
where
%%%(2.6, 2.7)
\begin{eqnarray}
G_{ij}(x,t) &=& g_i(x,t)g_j(x,t)^* \\
            &=& \frac{1}{2\pi}\int_{-\infty}^{\infty}dp_i\int_{-\infty}^{\infty}dp_j'f_i(p_i)f_j(p_j')^*e^{i[(p_i-p_j')x-(E_i-E_j')t]}~.
\end{eqnarray}

As $f_i(p)$, we take, for simplicity and definiteness,
%%%(2.8)
\begin{equation}
f_i(p_i) = \sqrt{\frac{1}{\sqrt{\pi}\Delta p_i}}\exp[-\frac{(p_i-p_i^0)^2}{2(\Delta p_i)^2}]~.
\end{equation}
Expanding $E_i(p_i)$ around $p_i^0$,\footnote{If one expands $E_i(p_i)$ as
\[
E_i \simeq E_i^0+(dE_i/dp_i)_{p_i=p_i^0}(p_i-p_i^0)+(d^2E_i/dp_i^2)_{p_i=p_i^0}(p_i-p_i^0)^2/2~,
\]
one may discuss spreading of the wave packets with time.\cite{rf:7} \ This is however beyond the scope of the present investigation.}
\[
E_i \simeq E_i^0+\beta_i^0(p_i-p_i^0) ~, 
\]
where
\[
\beta_i^0=(dE_i/dp_i)_{p_i=p_i^0}=p_i^0/E_i^0~,
\]
and performing integration over $p_i$, one obtains
%%%(2.9)
\begin{equation}
g_i(x,t) = \psi_i(x-\beta_i^0t)e^{i \theta_i^0(x,t)}~,
\end{equation}
where
%%%(2.10, 2.11)
\begin{eqnarray}
\psi_i(x) &=& \sqrt{\frac{1}{\sqrt{\pi}\Delta x_i}}\exp[-\frac{x^2}{2(\Delta x_i)^2}]~, \\
\theta_i^0(x,t) &=& p_i^0 x-E_i^0 t~, 
\end{eqnarray}
$E_i^0 = \sqrt{(p_i^0)^2+m_i^2}~$ and $\Delta x_i = 1/\Delta p_i~$. Substituting Eqs.(2.9) $\sim$ (2.11) into Eq.(2.6) and in turn into Eq.(2.5), one gets
%%%(2.12,2.13,2.14)
\begin{eqnarray}
G_{ij}(x,t) &=& \Psi_{ij}(x,t)e^{i\Theta_{ij}^0(x,t)}~, \\ \no \\
\Psi_{ij}(x,t) &=& \psi_i(x-\beta_i^0t)\psi_j(x-\beta_j^0t)~ \no \\
&=& \sqrt{\frac{1}{\pi\Delta x_i\Delta x_j}}\exp[-\frac{(x-\beta_i^0t)^2}{2(\Delta x_i)^2}-\frac{(x-\beta_j^0t)^2}{2(\Delta x_j)^2}] \no \\
&=& \sqrt{\frac{1}{\pi\Delta x_i\Delta x_j}}\exp[-\frac{(x-\beta_{ij}^0t)^2((\Delta x_i)^2+(\Delta x_j)^2)}{2(\Delta x_i)^2(\Delta x_j)^2} - \frac{(\beta_i^0-\beta_j^0)^2t^2}{2((\Delta x_i)^2+(\Delta x_j)^2)}]~, \no \\ \\
\Theta_{ij}^0(x,t) &=& \theta_i^0(x,t)-\theta_j^0(x,t) \no \\
                   &=& (p_i^0-p_j^0)x - (E_i^0-E_j^0)t~,
\end{eqnarray}
and\footnote{$U=(U_{\alpha i})$ is taken to be a real orthogonal matrix here.}
%%%(2.15)
\begin{eqnarray}
P_{\alpha \to \alpha'}(x,t) = \sum_iU_{\alpha i}^2U_{\alpha 'i}^2\Psi_{ii}(x,t) + 2\sum_{i<j}U_{\alpha i}U_{\alpha' i}U_{\alpha j}U_{\alpha' j}\Psi_{ij}(x,t)\cos\Theta_{ij}^0(x,t)~, \no \\
\end{eqnarray}
where
%%%(2.16)
\begin{eqnarray}
\beta_{ij}^0 &=& \frac{\beta_i^0(\Delta p_i)^2 + \beta_j^0(\Delta p_j)^2}{(\Delta p_i)^2 + (\Delta p_j)^2}~ \no \\
&=& \frac{\beta_i^0/(\Delta x_i)^2 + \beta_j^0/(\Delta x_j)^2}{1/(\Delta x_i)^2 + 1/(\Delta x_j)^2}~.
\end{eqnarray}
We shall hereafter refer to $\Psi_{ij}(x,t)$ and $\Theta_{ij}^0(x,t)$ as the wave-packet factor and the phase factor, respectively. \\
\ \\
\underline{Trajectory of the neutrino wave packets} \\

$\Psi_{ij}(x,t)$ as a function of $x$ has a peak at
%%%(2.17)
\begin{equation}
x = \beta_{ij}^0t~,
\end{equation}
and one has
%%%(2.18,2.19)
\begin{eqnarray}
\Psi_{ij}(\beta_{ij}^0t,t) &=& \sqrt{\frac{1}{\pi\Delta x_i\Delta x_j}}\exp[-\frac{(\beta_i^0-\beta_j^0)^2t^2}{2((\Delta x_i)^2+(\Delta x_j)^2)}]~, \\
\Theta_{ij}^0(\beta_{ij}^0t,t) &=& [(p_i^0-p_j^0)\beta_{ij} - (E_i^0-E_j^0)]t \no \\
&=& -[\frac{m_i^2-m_j^2}{E_i^0+E_j^0} + \frac{(\beta_i^0-\beta_j^0)(p_i^0-p_j^0)(E_i^0(\Delta x_i)^2-E_j^0(\Delta x_j)^2)}{(E_i^0+E_j^0)((\Delta x_i)^2+(\Delta x_j)^2)}]t~. \no \\
\end{eqnarray}
Eq.(2.18) tells us that $\Psi_{ij}(\beta_{ij}^0t,t)$ would reduce by a factor of $e$ or more compared to its value at $t=0$ unless 
%%%(2.20)
\begin{equation}
|(\beta_i^0-\beta_j^0)|t < \sqrt{2((\Delta x_i)^2+(\Delta x_j)^2)}~.
\end{equation}
This gives the coherence condition for neutrino oscillation.\cite{rf:3} \ Similarly, $\Psi_{ij}(x,t)$ regarded as a function of $t$ has a peak at
%%%(2.21)
\begin{equation}
t=x/\beta_{ij}^{0\prime}~, 
\end{equation}
with $\beta_{ij}^{0\prime}$ given by
%%%(2.22)
\begin{eqnarray}
\beta_{ij}^{0\prime} &=& \frac{(\beta_i^0)^2(\Delta p_i)^2 + (\beta_j^0)^2(\Delta p_j)^2}{\beta_i^0(\Delta p_i)^2 + \beta_j^0(\Delta p_j)^2} \no \\
&=& \frac{(\beta_i^0)^2/(\Delta x_i)^2 + (\beta_j^0)^2/(\Delta x_j)^2}{\beta_i^0/(\Delta x_i)^2 + \beta_j^0/(\Delta x_j)^2}~,
\end{eqnarray}
and one finds
%%%(2.23)
\begin{equation}
\Psi_{ij}(x,x/\beta_{ij}^{0\prime}) = \sqrt{\frac{1}{\pi\Delta x_i\Delta x_j}}\exp[-\frac{(1/\beta_i^0-1/\beta_j^0)^2x^2}{2((\Delta x_i/\beta_i^0)^2+(\Delta x_j/\beta_j^0)^2)}]~,
\end{equation}
%%%(2.24)
\begin{eqnarray}
\Theta_{ij}^0(x,x/\beta_{ij}^{0\prime}) &=& ~~[(p_i^0-p_j^0) - (E_i^0-E_j^0)/\beta_{ij}']x \no \\
&=& -[\frac{m_i^2-m_j^2}{p_i^0+p_j^0}-\frac{(1/\beta_i^0-1/\beta_j^0)(E_i^0-E_j^0)(p_i^0(\Delta x_i/\beta_i^0)^2-p_j^0(\Delta x_j/\beta_j^0)^2)}{(p_i^0+p_j^0)((\Delta x_i/\beta_i^0)^2+(\Delta x_j/\beta_j^0)^2)}] x~. \no \\
\end{eqnarray}
Eq.(2.23) gives another form of coherence conditions for neutrino oscillation:
%%%(2.25)
\begin{equation}
|(1/\beta_i^0-1/\beta_j^0)|x < \sqrt{2((\Delta x_i/\beta_i^0)^2+(\Delta x_j/\beta_j^0)^2)}~.
\end{equation}
\ \\
\underline{Space- and time-integration prescriptions} \\

Let us now calculate 
%%%(2.26)
\begin{equation}
G_{ij}(t) = \int_{-\infty}^{\infty}dx G_{ij}(x,t)~.
\end{equation}
Substituting Eqs.(2.12) $\sim$ (2.14), and performing integration over $x$, one obtains

%%%(2.27)
\begin{eqnarray}
G_{ij}(t) &=& \sqrt{\frac{2\Delta x_i\Delta x_j}{(\Delta x_i)^2+(\Delta x_j)^2}}~e^{i\Theta_{ij}^0(\beta_{ij}^0t,t)} \no \\
&\times& \exp[-\frac{(\beta_i^0-\beta_j^0)^2 t^2}{2((\Delta x_i)^2+(\Delta x_j)^2)}] \no \\
&\times& \exp[-\frac{(p_i^0-p_j^0)^2(\Delta x_i)^2(\Delta x_j)^2}{2((\Delta x_i)^2+(\Delta x_j)^2)}]~.
\end{eqnarray}
Note that the same phase factor as Eq.(2.19) appears here. In the right-hand side of Eq.(2.27), the second line gives the coherence condition (2.20), while the third line gives, when $p_i^0 \neq p_j^0$, another condition for neutrino oscillation to occur. These conditions, in the case of $\Delta p_i$ (or $\Delta x_i$) being independent of $i$, read
%%%(2.28)
\begin{equation}
|(\beta_i^0-\beta_j^0)|t/2 < \Delta x = 1/\Delta p < 2/|p_i^0-p_j^0|~.
\end{equation}

Calculating
%%%(2.29)
\begin{equation}
G_{ij}(x) = \int_{-\infty}^{\infty}dt G_{ij}(x,t)
\end{equation}
in a similar way, one obtains
%%%%(2.30)
\begin{eqnarray}
G_{ij}(x) &=& \sqrt{\frac{2(\Delta x_i/\beta_i^0)(\Delta x_j/\beta_j^0)}{\beta_i^0\beta_j^0((\Delta x_i/\beta_i^0)^2 + (\Delta x_j/\beta_j^0)^2)}}~e^{i\Theta_{ij}^0(x,x/\beta_{ij}^{0\prime})} \no \\
&\times& \exp[-\frac{(1/\beta_i^0-1/\beta_j^0)^2 x^2}{2((\Delta x_i/\beta_i^0)^2+(\Delta x_j/\beta_j^0)^2)}] \no \\
&\times& \exp[-\frac{(E_i^0-E_j^0)^2(\Delta x_i/\beta_i^0)^2(\Delta x_j/\beta_j^0)^2}{2((\Delta x_i/\beta_i^0)^2 + (\Delta x_j/\beta_j^0)^2)}]~,
\end{eqnarray}
where $\Theta_{ij}^0(x,x/\beta_{ij}^{0\prime})$ is given by Eq.(2.24). A result similar to Eq.(2.30) was obtained before by Giunti, Kim and Lee\cite{rf:7} for the case of $\Delta p_i$ (or $\Delta x_i$) being independent of $i$, who noted that, in the right-hand side of this equation, the second line gives the coherence condition Eq.(2.25), and the third line acts as another factor to suppress the oscillation probability.

\section{Comparison with the conventional approaches}
 
In the conventional approach, it is supposed that $|\nu_{\alpha}(x,t)\ket$ is given by
%%%(3.1)
\begin{equation}
|\nu_{\alpha}(x,t)\ket =\sum_iU_{\alpha i}|\nu_i\ket e^{i\theta_i^0(x,t)}~,
\end{equation}
and, accordingly, $P_{\alpha \rightarrow \alpha'}(x,t)$ is given by
%%%(3.2)
\begin{equation}
P_{\alpha \to \alpha'}(x,t) = \sum_iU_{\alpha i}^2U_{\alpha' i}^2 + 2\sum_{i<j}U_{\alpha i}U_{\alpha' i}U_{\alpha j}U_{\alpha' j}\cos\Theta_{ij}^0(x,t)~,
\end{equation}
where $\theta_i^0(x,t)$ and $\Theta_{ij}^0(x,t)$ are given by Eqs.(2.11) and (2.14). If one further assumes either $E_i^0=E_j^0$ (equal-energy prescription) or $p_i^0=p_j^0$ (equal-momentum prescription), one has
%%%(3.3)
\begin{equation}
\Theta_{ij}^0(x,t) = (p_i^0-p_j^0)x = -(m_i^2-m_j^2)x/(p_i^0+p_j^0)~,
\end{equation}
or
%%%(3.4)
\begin{equation}
\Theta_{ij}^0(x,t) =-(E_i^0-E_j^0)t = -(m_i^2-m_j^2)t/(E_i^0+E_j^0)~,
\end{equation}
implying that $P_{\alpha \to \alpha'}(x,t)$ will oscillate as a function of $x$ or $t$ with wave length $\ell_{ij}=2\pi|(p_i^0+p_j^0)/(m_i^2-m_j^2)|$ or period $\tau_{ij}=2\pi|(E_i^0+E_j^0)/(m_i^2-m_j^2)|$. Neutrino oscillation is usually 
discussed on the basis of Eqs.(3.1) $\sim$ (3.4) and it is often claimed 
that these two prescriptions give practically the same results for relativistic neutrinos.\footnote{In our previous paper, on arguing that these two prescriptions are better to be conceptually distinguished from each other, we have, in order to distinguish from the oscillation length $\ell_{ij}$, referred to $\tau_{ij}$ as oscillation period. As for distinction between these two prescriptions, see also the arguments given by Lipkin.\cite{rf:5}}

With the plane-wave expression (3.1), nothing can be said about the trajectory of the neutrinos. If, however, one arbitrarily assumes that the neutrinos are on the trajectory given by
%%%(3.5)
\begin{equation}
x=\bar{\beta}_{ij}t~, \qquad \bar{\beta}_{ij} = (p_i^0+p_j^0)/(E_i^0+E_j^0)~,
\end{equation}
one would then find expressions exactly same with Eqs.(3.3) and (3.4):
%%%(3.6)
\begin{eqnarray}
\Theta_{ij}^0(x,x/\overline{\beta}_{ij}) &=& -(m_i^2-m_j^2)x/(p_i^0+p_j^0)~, \no \\
\Theta_{ij}^0(\overline{\beta}_{ij}t,t) &=& -(m_i^2-m_j^2)t/(E_i^0+E_j^0)~.
\end{eqnarray}
We shall refer to the trajectory described by Eq.(3.5) as "classical trajectory"\cite{rf:6} and to such an approach as the "center-of-mass velocity" prescription.\footnote{In Ref.8, the authors considered a frame defined by the velocity $\overline{\beta}_{ij}$, i.e., the center-of-mass frame of the two components of the particle in question.}

A couple of comments are in order.

(1) In calculating $G_{ij}(x)$, Eq.(2.29), if one substitutes Eq.(2.7)
    and performs integration over $t$ first, one would have a
    $\delta$-function with its argument given by $E_i-E_j'$, which
    implies "equal energy" explicitly.\footnote{In our previous
    wave-packet treatment,\cite{rf:1} \ we have imposed the condition of
    equal energy by hand, as usually done in the case of the plane-wave
    approach, and thereby derived the conditions for neutrino
    oscillation to occur, i.e., the conditions corresponding to the
    inequalities (2.28). These conditions have been well known and
    derived mostly from more or less intuitive arguments.\cite{rf:3}\cite{rf:4}\cite{rf:5}}
    Therefore, one may in this sense regard the conventional
    equal-energy prescription reviewed here as a simplified version of
    the time-integration prescription considered in the preceding
    section and regard the second term in Eq.(2.24), in the case of
    $E_i^0 \neq E_j^0$, as a correction term to Eq.(3.3). Similar
    comments apply to the equal-momentum prescription mentioned here and
    the space-integration prescription given in the preceding section. 
 
(2) The trajectory of the neutrino wave packets given by Eqs.(2.17) and (2.16), or given by Eqs.(2.21) and (2.22), differs in general from the trajectory given by Eq.(3.5). $\beta_{ij}^0$ ($\beta_{ij}^{0\prime}$) given by Eq.(2.16) (Eq.(2.22)) would coincide with $\bar{\beta}_{ij}^0$ given by Eq.(3.5), and the phase factor given by Eq.(2.19) (Eq.(2.24)) would coincide with that given by Eq.(3.4) (Eq.(3.3)), if the central values and the dispersions of the momentum distribution functions of the neutrinos satisfy
%%%(3.6)
\begin{equation}
\Delta p_i/\Delta p_j = \sqrt{E_i^0/E_j^0} \qquad (~\beta_i^0\Delta p_i/\beta_j^0\Delta p_j = \sqrt{p_i^0/p_j^0}~)~. 
\end{equation}
We shall shortly see that these relations do not necessarily hold.
\section{Neutrinos from pion decays}
Now let us consider the case in which the neutrinos in question are created as a result of the decay $\pi^+ \rightarrow \mu^+ + \nu_{\alpha}~~(\alpha \equiv \mu)$ (with a branching fraction of 100\%), and suppose that a $\pi^+$ with momentum $p_{\pi}$ and energy $E_{\pi}=E_{\pi}(p_{\pi})=\sqrt{p_{\pi}^2+m_{\pi}^2}$, 
moving towards the $x$-direction, emits at $t=0$ a $\nu_{\alpha}$ towards the $x$-direction (and a $\mu^+$, with mass $m_{\mu}$, towards the opposite direction) in its rest frame. The momentum $p_i$ and energy $E_i$ of the $\nu_i$ component of the $\nu_{\alpha}$ are given by
%%%(4.1)
\begin{eqnarray}
p_i=\gamma_{\pi}(p_i^*+\beta_{\pi} E_i^*)=(E_{\pi}p_i^*+p_{\pi}E_i^*)/m_{\pi}~, \no \\
E_i=\gamma_{\pi}(E_i^*+\beta_{\pi} p_i^*)=(E_{\pi}E_i^*+p_{\pi}p_i^*)/m_{\pi}~, 
\end{eqnarray}
where $\beta_{\pi}=p_{\pi}/E_{\pi}$ and $\gamma_{\pi}=1/\sqrt{1-\beta_{\pi}^2}=E_{\pi}/m_{\pi}$ are the velocity and $\gamma$-factor of the $\pi^+$, and $p_i^*$ and $E_i^*$ are the momentum and energy of the $\nu_i$ in the rest frame of the $\pi^+$,
%%%(4.2)
\begin{eqnarray}
E_i^* &=& (m_{\pi}^2+m_i^2-m_{\mu}^2)/2m_{\pi}~, \no \\
p_i^* &=& \sqrt{(E_i^*)^2-m_i^2} \no \\
      &=& \sqrt{[m_{\pi}^2-(m_{\mu}+m_i)^2][m_{\pi}^2-(m_{\mu}-m_i)^2]}~/2m_{\pi}~.
\end{eqnarray}
$p_i$ and $E_i$ are functions of $p_{\pi}$ and one readily verifies
%%%(4.3)
\begin{equation}
dp_i/dp_{\pi} = E_i/E_{\pi}~, \qquad dE_i/dp_{\pi} = p_i/E_{\pi}~.
\end{equation}

It is natural to suppose that the momentum distribution $f_i(p_i)$ of $\nu_i$ is determined by the momentum distribution $f_{\pi}(p_{\pi})$ of the parent $\pi^+$ and to formulate this relation quantitatively as
%%%(4.4)
\begin{equation}
|f_i(p_i)|^2~dp_i = |f_{\pi}(p_{\pi})|^2~dp_{\pi}~.
\end{equation}
As $f_{\pi}(p_{\pi})$, we take again a Gaussian function:
%%%(4.5)
\begin{equation}
f_{\pi}(p_{\pi}) = \sqrt{\frac{1}{\sqrt{\pi}\Delta p_{\pi}}}\exp[-\frac{(p_{\pi}-p_{\pi}^0)^2}{2(\Delta p_{\pi})^2}]~.
\end{equation}
Noting that
\[
dp_i \simeq (E_i^0/E_{\pi}^0)dp_{\pi}~, \qquad p_i \simeq p_i^0 + (E_i^0/E_{\pi}^0)(p_{\pi}-p_{\pi}^0)~,
\]
where 
\[
p_i^0 = p_i(p_{\pi}^0)~, \qquad E_i^0 = E_i(p_{\pi}^0)~,
\]
one may use Eq.(4.4) to translate $f_{\pi}(p_{\pi})$ into $f_i(p_i)$. This leads one to Eq.(2.8), with $\Delta p_i$ given by
%%%(4.6)
\begin{equation}
\Delta p_i = (E_i^0/E_{\pi}^0)\Delta p_{\pi}~,
\end{equation}
which implies
%%%(4.7)
\begin{equation}
\Delta p_i/\Delta p_j = E_i^0/E_j^0~,
\end{equation}
to be compared with Eqs.(3.7).

To appreciate difference between Eq.(3.7) and Eq.(4.7), it is instructive to note that, for the present case, the wave-packet factor $\Psi_{ij}(x,t)$, E.(2.13), may be expressed as
%%%(4.8)
\begin{eqnarray}
\Psi_{ij}(x,t) &=& \sqrt{\frac{E_i^0E_j^0}{\pi(E_{\pi}^0\Delta x_{\pi})^2}}\exp[-\frac{(E_i^0)^2(x-\beta_i^0t)^2 + (E_j^0)^2(x-\beta_j^0t)^2)}{2(E_{\pi}^0\Delta x_{\pi})^2}]~ \no \\
 &=& \sqrt{\frac{E_i^0E_j^0}{\pi(E_{\pi}^0\Delta x_{\pi})^2}}\exp[-\frac{(\xi_{ij}^{(+)}(x,t))^2 + (\xi_{ij}^{(-)}(x,t))^2}{(2\Delta x_{\pi})^2}]~,
\end{eqnarray}
where 
%%%(4.9)
\begin{equation}
\xi_{ij}^{(\pm)}(x,t) = \{(E_i^0 \pm E_j^0)x-(p_i^0 \pm p_j^0)t\}/E_{\pi}^0~,
\end{equation}
and $\Delta x_{\pi} = 1/\Delta p_{\pi}~$, and that the velocity of the wave packets, $\beta_{ij}^0$ or $\beta_{ij}^{0\prime}$, may be expressed as
%%%(4.10,4.11)
\begin{eqnarray}
\beta_{ij}^0 &=& \frac{(E_i^0+E_j^0)(p_i^0+p_j^0)+(E_i^0-E_j^0)(p_i^0-p_j^0)}{(E_i^0+E_j^0)^2+(E_i^0-E_j^0)^2}~, \\
1/\beta_{ij}^{0\prime} &=& \frac{(E_i^0+E_j^0)(p_i^0+p_j^0)+(E_i^0-E_j^0)(p_i^0-p_j^0)}{(p_i^0+p_j^0)^2+(p_i^0-p_j^0)^2}~.
\end{eqnarray}
Since, for relativistic neutrinos, the second term is negligible compared to the first term in the numerator as well as in the denominator in each of Eqs.(4.10) and (4.11), we see that the center-of-mass velocity prescription is good enough to describe oscillations of relativistic neutrinos emitted from pions.
\section{Discussion}

In Ref.1, we have, as another possible prescription, proposed the equal-velocity 
prescription and pointed out that this prescription would also lead one exactly to Eqs.(3.3) and (3.4)\footnote{Note that $\beta_{ij}^0=\beta_{ij}^{0\prime}=\bar{\beta}_{ij}^0=\beta_i^0$, if $\beta_i^0=\beta_j^0$.}and that, if the condition of equal-velocity indeed prevails, neutrino oscillations would be free from the coherence conditions, Eqs.(2.20) and (2.25). We have also mentioned that whether and how experiments to be described by the equal-energy/-momentum/-velocity prescriptions could become realistic and feasible remain to be carefully examined and contrived with efforts. Here 
we like to mention that the equal-energy (-momentum) prescription seems appropriate to approximately describe oscillation experiments which involve something 
corresponding to time- (space-) integration or averaging.\footnote{The
equal-velocity case was later considered also by De Leo et
al..\cite{rf:9} \ Ahluwalia\cite{rf:10} suggested that the issue of equal-velocity versus equal-energy/-momentum can be quite  important and should be
testable for supernova neutrinos, while Giunti\cite{rf:11} 
suggested that, although the equal-energy/-momentum case could be realized
approximately in some physical processes, the equal-velocity case seems
very unlikely in any physical process.}

For neutrinos produced in pion decays or in any two-body decays, none of equal-energy, equal-momentum and equal-velocity holds.\footnote{A similar remark was 
raised explicitly by Winter before.\cite{rf:12}}Explicit kinematical considerations 
combined with our key postulate Eq.(4.4) lead us to Eq.(4.7) and hence to 
an explicit example which indicates that the center-of-mass velocity prescription is not necessarily applicable, or stated in a different way, the wave packets do not necessarily follow the classical trajectory described by Eq.(3.5). 
Although we have at the same time confirmed that this prescription is good enough to describe oscillations of relativistic neutrinos emitted from pions, possible deviation from a classical picture should be taken into account in general.

To conclude, we like to remark that the main points discussed in the preceding 
sections and summarized above are approximation-dependent, and, for comparison, we shall give another way of approximation in Appendix B.

\section*{Acknowledgements}

We are grateful to Professor S.~Kamefuchi and Professor T.~Ohshima for illuminating discussions and to the members of the Particle Physics Group at Nihon University for continuous encouragement. We would like also to thank Professor D.~V.~Ahluwalia, Professor C.~Giunti, Professor J.~Lowe and Professor Y.~Srivastava for comments to our previous work.

\appendix
\section{Remarks on normalization} %Empty argument \section{} yields `Appendix'. 

With Eq.(2.4), one has
%%%(A.1,A.2,A.3)
\begin{eqnarray}
\int_{-\infty}^{\infty}dx\bra \nu_{\alpha'}(x,t)|\nu_{\alpha}(x,t)\ket &=& \delta_{\alpha'\alpha}~, \\
\int_{-\infty}^{\infty}dxP_{\alpha \rightarrow \alpha'}(x,0) &=& \sum_{i,j}U_{\alpha i}U_{\alpha' i}^*U_{\alpha j}^*U_{\alpha' j} \sqrt{\frac{2\Delta x_i\Delta x_j}{(\Delta x_i)^2+(\Delta x_j)^2}} \no \\
&\times& \exp[-\frac{(p_i^0-p_j^0)^2(\Delta x_i)^2(\Delta x_j)^2}{2((\Delta x_i)^2+(\Delta x_j)^2)}]~, \\
\sum_{\alpha'}\int_{-\infty}^{\infty}dxP_{\alpha \rightarrow \alpha'}(x,t) &=& 
\int_{-\infty}^{\infty}dx\bra \nu_{\alpha}(x,t)|\nu_{\alpha}(x,t)\ket \no \\
&=& 1~.
\end{eqnarray}
Eq.(A.2) implies that, with our normalization (2.4), the initial condition stated in the beginning of Sect.2 as "suppose a neutrino of flavor $\alpha$ is born at $t=0$" is realized strictly only in the limit of $\Delta x_i=\Delta x_j \rightarrow 0$ or if the momentum distribution functions $f_i(p_i)$ have a common central value and a common dispersion.

If, as a normalization condition alternative to Eq.(2.4), one adopts
%%%(A.4)
\begin{equation}
\sqrt{\frac{1}{2\pi}}\int_{-\infty}^{\infty}dp_if_i(p_i) = 1~,
\end{equation}
one would have
%%%(A.5,A.6,A.7)
\begin{eqnarray}
\bra \nu_{\alpha'}(x,t)|\nu_{\alpha}(x,t)\ket &=& \delta_{\alpha'\alpha}~, \\
P_{\alpha \rightarrow \alpha'}(0,0) &=& \delta_{\alpha'\alpha}~, \\
\sum_{\alpha'}P_{\alpha \rightarrow \alpha'}(x,t) &=& \bra \nu_{\alpha}(x,t)|\nu_{\alpha}(x,t)\ket \no \\
&=& \sum_i U_{\alpha i}U_{\alpha i}^*\exp[-(x-\beta_i^0t)^2/(\Delta x_i)^2]~.
\end{eqnarray}
This way of normalization has advantage that it keeps the correspondence
with the conventional plane-wave treatment as far and transparent as
possible. In fact, this normalization ensures Eqs.(2.2) and (2.5) to
reduce respectively to Eqs.(3.1) and (3.2), and, accordingly,
$\sum_{\alpha'}P_{\alpha \rightarrow \alpha'}(x,t) = \bra
\nu_{\alpha}(x,t)|\nu_{\alpha}(x,t)\ket \rightarrow 1$, when $\Delta p_i
=1/\Delta x_i \rightarrow 0$. Although the main points we discussed in
the text are conceivably independent of how $f_i(p_i)$ is normalized,
the problem of normalization itself involves some subtlety and ambiguity
and deserves to be studied further.\footnote{Eq.(A.4) was adopted in our
first paper,\cite{rf:1} \ and Eq.(2.4) has been adopted in our subsequent
discussions.\cite{rf:13} \ See Ref.7 for another normalization procedure.} 

\section{Another way of approximation} %Empty argument \section{} yields `Appendix'. 

To calculate $G_{ij}(x,t)$ (defined by Eq.(2.7)) for $i \neq j$, if one approximates the exponent $(p_i-p_j')x-(E_i-E_j')t$ as
%%%(B.1)
\begin{eqnarray}
(p_i-p_j')x &-& (E_i-E_j')t \no \\
&=& (p_i-p_j')x-[(p_i^2+m_i^2)-(p_j^{\prime 2}+m_j^2)]t/(E_i+E_j') \no \\
&=& (p_i-p_j')[x-(p_i+p_j')t/(E_i+E_j')]-(m_i^2-m_j^2)t/(E_i+E_j') \no \\
&\simeq& (p_i-p_j')[x-(p_i^0+p_j^0)t/(E_i^0+E_j^0)]-(m_i^2-m_j^2)t/(E_i^0+E_j^0) \no \\
&=& (p_i-p_j'-p_i^0+p_j^0)(x-\bar{\beta}_{ij}^0t) + \Theta_{ij}^0(x,t)~,
\end{eqnarray}
and performs integration, by changing the integration variables from $p_i$ and $p_j'$ to
\[
p_{ij}=p_i-p_j'~, \qquad P_{ij}=(p_i(\Delta p_j)^2+p_j'(\Delta p_i)^2)/((\Delta p_i)^2+(\Delta p_j)^2)~,
\]
one would be led to
%%%(B.2)
\begin{equation}
G_{ij}(x,t) = \sqrt{\frac{1}{\sqrt{\pi}\Delta x_i\Delta x_j}}e^{i\Theta_{ij}^0(x,t)}~\exp[-\frac{(x-\bar{\beta}_{ij}^0t)^2((\Delta x_i)^2+(\Delta x_j)^2)}{2(\Delta x_i)^2(\Delta x_j)^2}]~,
\end{equation}
a result compatible with the statement that the wave packets are on the classical trajectory, (3.5).\footnote{In cotrast, in Ref.6, the authors presupposed that the neutrinos should be on their classical trajectory, and treated the phase facor $(p_i-p_j')-(E_i-E_j')$ as
\begin{equation}
(p_i-p_j')x-(E_i-E_j')t~\simeq~-(m_i^2-m_j^2)t/(E_i^0+E_j^0)~,
%%%(B.3)
\end{equation}
which is to be compared with Eq.(B.1).}Note that Eq.(B.1) is to be compared with our approximation
%%%(B.4)
\begin{eqnarray}
(p_i-p_j')x&-&(E_i-E_j')t \no \\
&\simeq& (p_i-p_j')x-[E_i^0+\beta_i^0(p_i-p_i^0)-E_j^0-\beta_j^0(p_j'-p_j^0)]t \no \\
&=& (p_{ij}-p_{ij}^0)(x-\beta_{ij}^0t) - (P_{ij}-P_{ij}^0)(\beta_i^0-\beta_j^0)t + \Theta_{ij}^0(x,t)~,\end{eqnarray}
and Eq.(B.2) is to be compared with Eqs.(2.12) $\sim$ (2.14), where 
\[
p_{ij}^0=p_i^0-p_j^0~, \qquad P_{ij}^0=(p_i^0(\Delta p_j)^2+p_j^0(\Delta p_i)^2)/((\Delta p_i)^2+(\Delta p_j)^2)~,
\]
and $\beta_{ij}^0$ is defined by Eq.(2.16).


\begin{thebibliography}{99}
%%%%%%%%%%%%%%%%%%%%%%%%%%%%%%%%%%%%%%%%%%%%%%%%%%%%%%%%%%%%%
% Some macros are available for the bibliography:
%   o for general use
%      \JL : general journals          \andvol : Vol (Year) Page
%   o for individual journal 
%      \PR  : Phys. Rev.               \PRL : Phys. Rev. Lett.
%      \NP  : Nucl. Phys.              \PL  : Phys. Lett.
%      \JMP : J. Math. Phys.           \CMP : Commun. Math. Phys.
%      \PTP : Prog. Theor. Phys.       \JPSJ: J. Phys. Soc. Jpn.
%      \JP  : J. of Phys.              \NC  : Nouvo Cim.
%      \IJMP: Int. J. Mod. Phys.       \ANN : Ann. of Phys.
% Usage:
%   \PR{D45,1990,345}            ==> Phys.~Rev.\ {\bf D45} (1990), 345
%   \JL{Phys.~Lett.,A30,1981,56} ==> Phys.~Lett.\ {\bf A30} (1981), 56
%   \andvol{B123,1995,1020}      ==> {\bf B123} (1995), 1020
%%%%%%%%%%%%%%%%%%%%%%%%%%%%%%%%%%%%%%%%%%%%%%%%%%%%%%%%%%%%%

\bibitem{rf:1}
Y.~Takeuchi, Y.~Tazaki, S. Y.~Tsai and T.~Yamazaki, Mod. Phys.Lett. {\bf A14}, (1999),2329 ; Preprint(hep-ph/9809558). \\
S. Y.~Tsai, "Quantum Interference -From Neutrino Oscillation to $CP$ Violation in the $B^0-\overline{B^0}$ system-", in {\it Proc. of the 8'th B-Physics International Workshop}, Kawatabi, Miyagi, Japan, October 29-31, 1998, eds. K.~Abe et al., p.95.
\bibitem{rf:2}
B.~Pontecorvo, Zh. Eksp. Teor. Fiz. {\bf 33}, (1957), 549; ibid. {\bf
	34}, (1958), 247. \\
Z.~Maki, M.~Nakagawa and S.~Sakata, Prog. Theor. Phys. {\bf 28}, (1962), 870. \\
S.~M.~Bilenky and S.~T.~Petcov, Rev. Mod. Phys. {\bf 59}, (1987), 671.
\bibitem{rf:3}
S.~Nussinov, Phys. Lett. {\bf B63}, (1976), 201.
\bibitem{rf:4}
B.~Kayser, Phys. Rev. {\bf D24}, (1981), 110.
\bibitem{rf:5}
H.~J.~Lipkin, Phys. Lett. {\bf B348}, (1995), 604.
\bibitem{rf:6}
A.~D.~Dolgov, A.~Yu.~Morozov, L.~B.~Okun and M.~G.~Schepkin,
	Nucl. Phys. {\bf B502}, (1997), 3.
\bibitem{rf:7}
C.~Giunti, C.~W.~Kim and U.~W.~Lee, Phys. Rev. {\bf D44}, (1991), 3635.
\bibitem{rf:8}
H.~Burkhardt, J.~Lowe, G.~J.~Stephenson Jr. and T.~Goldman, Preprint (16-Mar-98).
\bibitem{rf:9}
S.~De Leo, G.~Ducati and P.~Rotelli, Preprint (hep-ph/9906460).
\bibitem{rf:10}
D.~V.~Ahluwalia, private communication.
\bibitem{rf:11}
C.~Giunti, private communication.
\bibitem{rf:12}
R.~G.~Winter, Lett. Nuovo Cim. {\bf 30}, (1981), 101.
\bibitem{rf:13}
	T.~Yamazaki, master thesis (Nihon University, March 1999, in Japanese). \\
K.~Matsuda, S.~Suzuki, Y.~Takeuchi and S.~Y.~Tsai, talk presentd by Matsuda at the Spring Meeting of Physical Society of Japan (Kinki University, Osaka, March 31,2000).


\end{thebibliography}
\end{document}